\documentclass[journal]{IEEEtran}
\ifCLASSINFOpdf
\else
\fi
\usepackage{amsmath,amssymb}
\usepackage{mathtools}
\DeclarePairedDelimiter{\ceil}{\lceil}{\rceil}
\usepackage{graphicx}
\usepackage{graphicx}
\usepackage{xcolor}
\usepackage[bottom]{footmisc}

\usepackage{epstopdf}
\usepackage{algorithm}
\usepackage[noend]{algpseudocode}
\usepackage{etoolbox}
\makeatletter
 \patchcmd{\@maketitle}
  {\addvspace{0.5\baselineskip}\egroup}
  {\addvspace{-1.5\baselineskip}\egroup}
  {}
  {}
\makeatother
\begin{document}
\raggedbottom

%
\title{An Energy-Efficient Transaction Model for the Blockchain-enabled Internet of Vehicles (IoV)}
%
%
%

\author{Vishal Sharma,~\IEEEmembership{Member,~IEEE,}
\thanks{V. Sharma is with the Department of Information Security Engineering, Soonchunhyang University, Asan-si 31538, Republic of Korea, Email: vishal\_sharma2012@hotmail.com.}
}

%
%

\markboth{IEEE Communications Letters}%
{}
%



\maketitle
\begin{abstract}
The blockchain is a safe, reliable and innovative mechanism for managing numerous vehicles seeking connectivity. However, following the principles of the blockchain, the number of transactions required to update ledgers pose serious issues for vehicles as these may consume the maximum available energy. To resolve this, an efficient model is presented in this letter which is capable of handling the energy demands of the blockchain-enabled Internet of Vehicles (IoV) by optimally controlling the number of transactions through distributed clustering. Numerical results suggest that the proposed approach is 40.16\% better in terms of energy conservation and 82.06\% better in terms of the number of transactions required to share the entire blockchain-data compared with the traditional blockchain.
\end{abstract}


%
\IEEEpeerreviewmaketitle

\section{Introduction}
Blockchain enables shared access to information which is broadcasted across a network based on the trust of its participants~\cite{gu2018consortium}. Industries aiming at services for the Internet of Vehicles (IoV) consider blockchain as a leading technology for handling managerial as well as transmission aspects of vehicles~\cite{liu2018blockchain}. Research groups in Mobility Open Blockchain Initiative (MOBI) have highlighted the potential use of this technology for IoV~\cite{uhlemann2018time}. Challenges like broadcast collision-avoidance, resource scheduling~\cite{wei2018qos}, and privacy-preserving~\cite{lim2017preserving} in IoV can be resolved through an efficient implementation of vehicular-blockchain. Blockchain enables IoV to be protected against different types of cyber threats as well as allows secure distribution of vehicular services. In spite of playing a pivotal role, there are certain issues related to the prominent use of this blockchain technology, which includes excessive energy consumption for ledger-updates, peer-to-peer and peer-to-multi-peer smart contracts, and excessive transactions~\cite{8326513}. Resolving these can help to extend the utility as well as the applicability of blockchain to IoV.

In the networks where communications are bounded by blockchain technology, the number of transactions on each node is considerably high and these transactions cause a huge impact on the consumption of energy resources. Moreover, the mechanisms in blockchain use public key operations, which may cause considerable overheads even if light-weight cryptography algorithms are used for onboard security. However, the primary cause of concern is the multiple exchanges of transaction-messages between the vehicles and the core, which is a problem and it is desired to find a solution to reduce these number of transactions. 
\begin{figure}[!ht]
  \centering
  \includegraphics[width=180px]{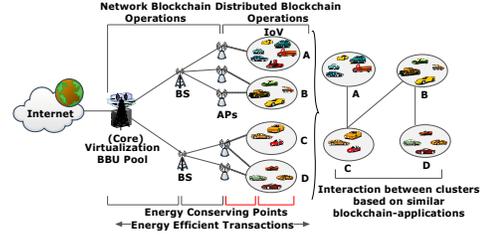}
  \caption{An exemplary illustration of IoV with energy conserving points.}\label{fig1}
\end{figure}

\section{System Model and Problem Statement}
The solution presented in this letter uses distributed clustering mechanism based on stochastic volatility model of security derivatives for reducing the burden of transactions on each device in IoV by finding the optimal slots for updating the blockchain ledgers. In general, the presented optimal transaction model is driven by the selection of appropriate Cluster-Heads (CHs) as a part of distributed clustering. The success of the approach depends on the reduction in the overall cost of operation, which is expressed in terms of energy-efficiency and the number of iterations performed to identify the final transaction in the formulated blockchain.

To further clarify, the problem at hand can be observed from Fig.~\ref{fig1} in which each entity plays the role of a miner where consensus is performed through broadcasting. Let $\mathcal{I}$ denote the set of vehicles each operating a set $\mathcal{S}$ of applications that require blockchain procedures to accomplish their operations, such that the total energy consumed by $i$th vehicle for $|\mathcal{S}|$ number of blockchains is expressed as:
\begin{equation}\label{eq:1}
\fontsize{8}{9}\selectfont
  \mathcal{B}^{(t)}_{\mathcal{S}, i}=\sum_{i=1}^{|\mathcal{S}|} \left(\beta_{C,M}^{(t)} +\left(\beta_{R}^{(t)}+\beta_{U}^{(t)}\right)\right)_{i},
\end{equation}
where $\beta_{C,M}^{(t)}$ is the energy consumed for security operations such as light-weight elliptic curve cryptography for checking correctness of nodes, $\beta_{R}^{(t)}$ and $\beta_{U}^{(t)}$ are the energy requirements of transmission procedures and blockchain-update (Ledgers) operations, respectively. In the proposed model,
\begin{equation}\label{eq:2}
\fontsize{8}{9}\selectfont
  \beta_{U}^{(t)}=\mathcal{H}\left(\mathcal{R}_{C}\times \mathcal{E}_{R}\right),
\end{equation}
and
\begin{equation}\label{eq:3}
\fontsize{8}{9}\selectfont
  \beta_{R}^{(t)}=\mathcal{H}\left(\sum_{j=1}^{k} \left(\mathcal{E}_{C}\times \gamma\right)_{j}\right),
\end{equation}
where $\mathcal{H}$ is the number of intermediate hops, $\mathcal{R}_{C}$ is the number of records updated in a transaction, $\mathcal{E}_{R}$ is the per record energy consumption, $\mathcal{E}_{C}$ is the per request energy consumption, $\gamma$ is the number of requests and $k$ is the types of messages like send, receive and acknowledgement. The network conditions for optimality are based on governing equations which are derived through a stochastic volatility model as it helps to predict the variation in energy consumption of the network. It also allows to accurately calculate the network sustaining rate if no additional batteries/resources are provided to IoV, such that the network differential variation, by using Heston Model~\cite{benhamou2010time}, is given as:
\begin{equation}\label{eq:4}
\fontsize{8}{9}\selectfont
  \frac{d\mathcal{B}}{dt}=\lambda\left(\mathcal{B}^{(t)}_{\mathcal{S}, i}- \mathcal{B}^{(o)}_{\mathcal{S}, i}\right)+\epsilon \sqrt{\mathcal{\sigma}} \frac{d\mathcal{B}'}{dt},
\end{equation}
where $\mathcal{B}^{(o)}_{\mathcal{S}, i}$ is the initial values for energy consumption, $\lambda$ is the number of blockchain requests per unit time, $\epsilon$ is the ratio of the excessive energy to the total initial energy of the vehicle, $\sigma$ is the standard deviation of energy required w.r.t. energy utilized, and $\frac{d\mathcal{B}'}{dt}$ denotes the rate of change of requests per unit time. Based on the given model, the location of a vehicle can be marked as $\mathcal{L}=f(x,y)$, and the maximum distance to connect can be written as $\mathcal{R}$, which shows the range between entities involved in the update and sharing of ledgers. To sustain at a given energy rate, the network can be modeled into a transfer function $\mathcal{F}_{t}$, which is given as:
\begin{equation}\label{eq:5}
\fontsize{8}{9}\selectfont
  \mathcal{F}_{t}^{(\tau)}=\sum\limits_{i=1}^{|\mathcal{C}|}\sum\limits_{j=1}^{|\mathcal{I}|}\sum\limits_{k=1}^{|\mathcal{S}|} \left(\mathcal{P}_{c}^{(A)} \times \mathcal{P}(f(p))\right),
\end{equation}
where $\mathcal{P}_{c}^{(A)}$ is the probability denoting the presence of a receiver, $\mathcal{P}(f(p))$ denotes the probability function that a vehicle is in the range, $\mathcal{C}$ is the set of clusters, $\tau$ is the operational time, and
\begin{equation}\label{eq:6}
\fontsize{8}{9}\selectfont
  \mathcal{P}(f(p))=1-\int_{0}^{\mathcal{R}} \mathcal{P}_{c} \times f(p) dp,
\end{equation}
where $f(p)$ is the vehicle movement function within $\mathcal{R}$ and $P_{c}$ is the probability of existence of vehicles with the given movement function ($f(p)$). Now, based on the above conditions, the network problem deals with attaining maximum transfers at a low consumption of energy, i.e.,
\begin{equation}\label{eq:7}
\fontsize{8}{9}\selectfont
\max\left(\mathcal{F}_{t}^{(\tau)}\right), \forall\;\mathcal{C},\;\forall\mathcal{I},\;\forall\mathcal{S},
\end{equation}
s.t.
\begingroup\makeatletter\def\f@size{7}\check@mathfonts$$
\min\left(\frac{d\mathcal{B}}{dt}\right), \forall\;\mathcal{C},\;\forall\mathcal{I},\;\forall\mathcal{S},$$\endgroup
\begingroup\makeatletter\def\f@size{7}\check@mathfonts$$\gamma \leq \frac{\tau'}{\tau}, \underbrace{\tau'}_{\text{vehicle stay time}} \leq \tau,$$\endgroup
\begingroup\makeatletter\def\f@size{7}\check@mathfonts$$1-\int_{0}^{\mathcal{R}} \mathcal{P}_{c} \times f(p) dp \;\;\geq \underbrace{\mathcal{P}_{th}}_{\text{threshold Probability}},$$\endgroup
\begingroup\makeatletter\def\f@size{7}\check@mathfonts$$\min\left(\int_{0}^{\tau}\int_{0}^{\mathcal{R}}\mathcal{P}(f(p)).\underbrace{\epsilon(t)}_{\text{error in transfer}} dp \;dt \right) \forall\;\mathcal{C},\;\forall\mathcal{I},\;\forall\mathcal{S}.~\refstepcounter{equation}(\theequation)\label{eq:8}$$\endgroup
\section{Distributed clustering for Blockchain-enabled IoV}
Usually, clustering involves a centralized entity, which is against the principles of the blockchain. However, with a distributed phenomenon over the selection of CHs, the principles of blockchain remain intact and the network can be managed efficiently. Distributed clustering helps to lower the number of updates for ledgers as well as reduces the number of blockchain-links ($\psi$) generated for each ledger operation. Moreover, the slot-wise transactions offer better control over the operations of the entire network. However, it is required that the network entities must be aware of the optimal points as well as know the slots, which may result in redundant transmissions. In such a way, the energy of the network can be conserved considerably. In general, the major factors in the proposed approach are about the selection of CHs, decision to transmit (when and how), the number of permissible blockchain queries, and location-based ledger-offloading. Satisfaction of all these issues through stochastic volatility helps to sustain the network for longer durations.
\begin{figure}[!ht]
  \centering
  \includegraphics[width=180px]{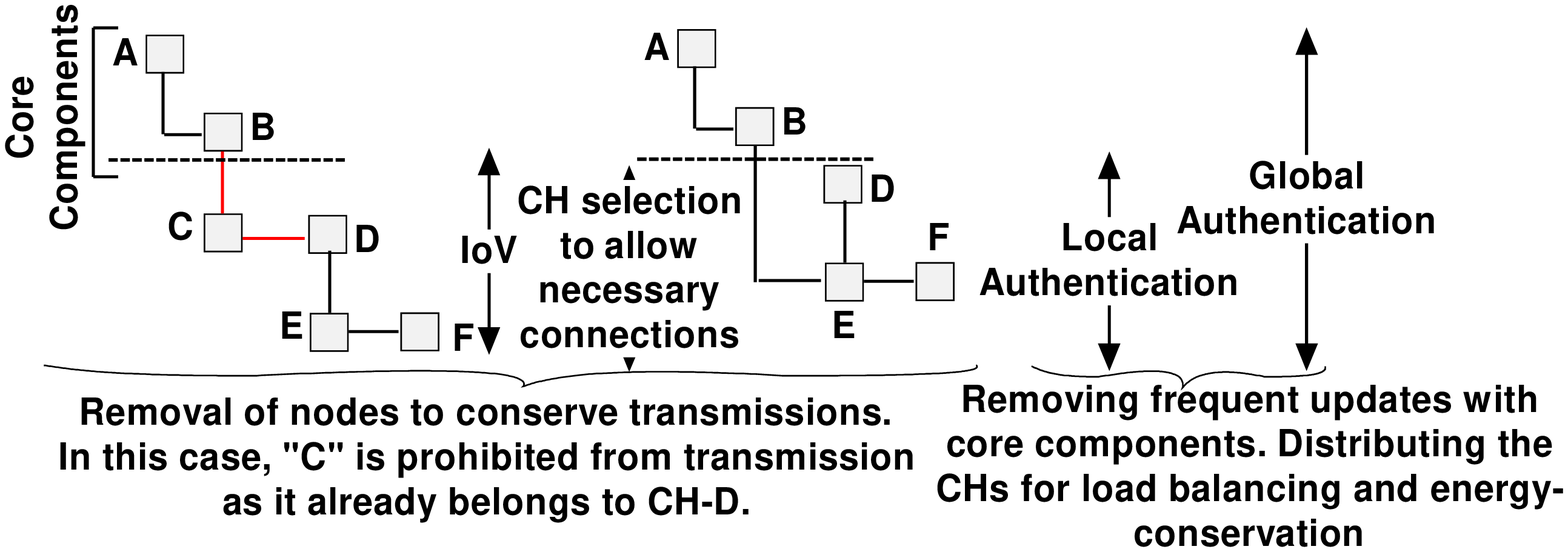}
  \caption{An exemplary illustration of the proposed mechanism of distributed clustering for the blockchain-enabled IoV.}\label{fig2}
\end{figure}
The mechanism of distributed clustering can be followed from Fig.~\ref{fig2}, according to which, every chain is labeled sequentially to identify the nodes involved in updates. The regular operations and authentication procedures are the same as defined in the original blockchain. However, in the proposed model, local and global chains are used for conserving energy, in which the ones with critical energy values (marked in red) are removed from the global operations and an appropriate CH is selected to handle their operations. For this, the local and global authentication processes are used, which involves hash-based authentication of CHs with the core components and hash-based authentication of general vehicles with CH. The CHs are temporary and their choices are made on-demand and as per the state of the network. This mechanism not only reduces the burden of excessive sharing but also lowers the complexity involved in performing fork operations for individual entries. Following Lemmas help to understand the operational details of the proposed methodology.\\
\textbf{Lemma-1} In the given model, the sustaining rate of the network is governed by $f(p)$ and $\lambda$, and the capacity of the network decreases if $\lambda$ increases to a large extent. In such a case, if the rate of change in the number of requests approaches zero, the lifetime becomes only a function of initial value of energy and per-unit-energy consumption of the blockchain application. \\
\textbf{Proof:} A blockchain network should limit to share the operations for maintaining network secrecy. However, the success of the proposed approach depends on the knowledge of the operations. According to which, the network governing equations can be calculated based on two frequencies, $f_{1}$ and $f_{2}$, of vehicle slots for operations defined in (\ref{eq:1}). Considering that the vehicles operate in a Gaussian mode, the two frequencies for network and blockchain operations can be given as $f_{1}=\frac{1}{\sigma_{1}\sqrt{2\pi}} e^{\frac{-1}{2} \left( \frac{\lambda_{1}-\overline{\lambda_{1}}}{\sigma_{1}}\right)^{2}}$ and $f_{2}=\frac{1}{\sigma_{2}\sqrt{2\pi}} e^{\frac{-1}{2} \left( \frac{\lambda_{2}-\overline{\lambda_{2}}}{\sigma_{2}}\right)^{2}}$, respectively. Gaussian distribution suites these formations as the average mean for network is high. Similar distribution can be used for f(p) in (\ref{eq:5}). Here, $\sigma_1$ and $\sigma_2$ are the standard deviations for general operations per unit time and blockchain operations per unit time, respectively. Now, using $\beta^{(t)}_{\mathcal{D},i} =\frac{\mathcal{B}^{(o)}_{\mathcal{S}, i}}{|\mathcal{S}|}\int_{0}^{\tau} e^{-(\lambda_{1}+\lambda_{2})t} dt$ and $\lambda_{k}=\sigma_{k}\sqrt{\ln(\sqrt{2\pi}\sigma_{k}f_{k})^{-2}}+\overline{\lambda_{k}},\; k=1,2,\; \lambda=\lambda_1+\lambda_2$, with $\overline{\lambda_{1}}\neq \overline{\lambda_{2}}$, and $f_{1}\neq f_{2}$, the energy decaying in the network for each vehicle can be observed by solving the integral ($\beta^{(t)}_{\mathcal{D},i}$), which gives
\begingroup\makeatletter\def\f@size{7}\check@mathfonts$$\frac{\mathcal{B}^{(o)}_{\mathcal{S}, i}}{|\mathcal{S}|}.\dfrac{1-\left(\mathrm{e}^{-\left(\sqrt{2}\left(\sigma_2\sqrt{-\ln\left(\sqrt{2 \pi}f_2\sigma_2\right)}+\sigma_1\sqrt{-\ln\left(\sqrt{2 \pi}f_1\sigma_1\right)}\right)+\overline{\lambda_2}+\overline{\lambda_1}\right)\tau}\right)}{\sqrt{2}\left(\sigma_2\sqrt{-\ln\left(\sqrt{2 \pi}f_2\sigma_2\right)}+\sigma_1\sqrt{-\ln\left(\sqrt{2 \pi}f_1\sigma_1\right)}\right)+\overline{\lambda_2}+\overline{\lambda_1}},~\refstepcounter{equation}(\theequation)\label{myeq1}$$\endgroup
as an output. In actual practice, $\lambda_{2}$ is unknown, however, its value can be estimated by assuming that all the operations in the network are synchronized, such that $\overline{\lambda_{1}}=\overline{\lambda_{2}}$ and $f_{1}= f_{2}$. The deviation is varied to maintain a difference in regular and blockchain operations. Now, in such a case, $f_{2}=f_{1}=\frac{1}{2\pi \sigma_{1}\sigma_2} e^{\frac{-1}{2}((\frac{\lambda_{1}-\overline{\lambda_1}}{\sigma_{1}})^{2})+(\frac{\lambda_{2}-\overline{\lambda_2}}{\sigma_{2}})^{2})}$ and $\lambda_{2}$ becomes $\sigma_2 \sqrt{\ln\left(2\pi\sigma_1 \sigma_2 f_1 \right)^{-2}-\left(\frac{\lambda_{1}-\overline{\lambda_1}}{\sigma_1}\right)^{2}}+\overline{\lambda_{1}}$, based on which, the energy decaying is calculated as \begingroup\makeatletter\def\f@size{7}\check@mathfonts$$\frac{\mathcal{B}^{(o)}_{\mathcal{S}, i}}{|\mathcal{S}|}.\dfrac{\sigma_1-\sigma_1\mathrm{e}^{-\frac{2\sigma_2\sqrt{-2\sigma_1^2\ln\left(2{\pi}f_1\sigma_1\sigma_2\right)-\overline{\lambda_1}^2+2\lambda_1\overline{\lambda_1}-\lambda_1^2}\tau}{\sigma_1}-2\overline{\lambda_1}\tau}}{2\left(\sigma_2\sqrt{-2\sigma_1^2\ln\left(2{\pi}f_1\sigma_1\sigma_2\right)-\overline{\lambda_1}^2+2\lambda_1\overline{\lambda_1}-\lambda_1^2}+\overline{\lambda_1}\sigma_1\right)}.~\refstepcounter{equation}(\theequation)\label{myeq2}$$\endgroup The values for these help to form an estimated overview of energy requirements/consumptions of each vehicle, and allow tracking of differential variations for energy demands in (\ref{eq:4}). Such provisioning facilitates the selection of CHs with a high availability of energy. In addition, the extent up to which the network can survive at a given depletion rate is observable through these results.\\
\textbf{Lemma-2} The number of transactions needed to shift the entire load depends on the number of local as well as global updates performed in the IoV. The local updates can be accumulated and transferred once a slot for global exchange is initiated. Based on these, the actual number of required transactions ($\mathcal{K}_{\mathcal{R}}^{(\mathcal{T})}$) is given by \begingroup\makeatletter\def\f@size{7}\check@mathfonts$$\ceil[\Bigg]{-\dfrac{\left(\sum_{i=1}^{|\mathcal{C}|}\sum_{j=1}^{\psi}\lambda\right)\mathcal{P}_{c}\tau^2\left(\operatorname{erf}\left(\mathcal{R'}-\mathcal{R''}\right)-\operatorname{erf}\left(\mathcal{R'}\right)\right)}{2^\frac{5}{2}D\sigma_{R''}}},~\refstepcounter{equation}(\theequation)\label{myeq3}$$\endgroup $\mathcal{R'}>0,\;\mathcal{R''}>0$ and $\tau > 0$ assuming that all vehicles follow similar pattern where $\mathcal{R''}$ is their radio range, $\mathcal{R'}$ is the average range of the network required to maintain a connection, $\sigma_{R''}$ is the deviation of $\mathcal{R''}$ from $\mathcal{R}$.\\
\textbf{Proof:} In the given network, the number of transactions are driven by $\lambda$, number of parallel links $D$, and the connectivity function of the network $\mathcal{G}(t)$. Here, $\mathcal{G}(t)$ is driven by $f(p)$ and $\mathcal{P}_{c}$, such that, the number of transitions to be performed for the blockchain can be written as $\frac{1}{D} \int_{0}^{\tau}\left(\int_{0}^{\mathcal{R''}} f(p).\mathcal{P}_{c}dp\right)\left(\sum_{i=0}^{|\mathcal{C}|}\sum_{i=0}^{\psi} \lambda.t \right) dt$, which on following the Gaussian distribution for vehicles can be written as $\frac{1}{D} \int_{0}^{\tau}\left(\int_{0}^{\mathcal{R''}} \frac{1}{\sqrt{2\pi}} e^{\frac{-1}{2}\left(\frac{x-\mathcal{R'}}{\sigma_{R''}} \right)^{2}}.\mathcal{P}_{c}\;dx\right)\left(\sum_{i=0}^{|\mathcal{C}|}\sum_{i=0}^{\psi} \lambda.t \right) dt$. Now, solving in parts, it can be written as \begingroup\makeatletter\def\f@size{7}\check@mathfonts$$\frac{1}{D} \left( \frac{-\mathcal{P}_{c} \left(erf\left(\frac{\mathcal{R'}-\mathcal{R''}}{\sqrt{2}\sigma_{R''}}\right)-erf\left(\frac{\mathcal{R'}}{\sqrt{2}\sigma_{R''}}\right) \right)}{2}\right) \int_{0}^{\tau}\left(\sum_{i=0}^{|\mathcal{C}|}\sum_{i=0}^{\psi} \lambda.t \right) dt.$$\endgroup By solving the remaining part, the observed value becomes $\ceil[\Bigg]{-\dfrac{\left(\sum_{i=1}^{|\mathcal{C}|}\sum_{j=1}^{\psi}\lambda\right)\mathcal{P}_{c}\tau^2\left(\operatorname{erf}\left(\mathcal{R'}-\mathcal{R''}\right)-\operatorname{erf}\left(\mathcal{R'}\right)\right)}{2^\frac{5}{2}D\sigma_{R''}}}$, which is the desired output at $\mathcal{R'}>0,\;\mathcal{R''}>0$ and $\tau > 0$. \\
\textbf{Optimal Solution:} The solution to the energy-efficient transition considering the distributed clustering mechanism can be attained by performing optimal offloading and selecting new CHs whenever the energy demand of the network increases. If $\mathcal{S}_{t,i}^{(\mathcal{I})}$ is the observation, which controls the changing of CH to optimize the transactions for conserving energy, the solutions can be attained through following formulations:
\begin{itemize}
  \item Based on Optimal Stopping Theory (OST)~\cite{Peskir2006}, change CH if
  \begin{equation}\label{eq:c1}
  \fontsize{7}{8}\selectfont
\max(\mathcal{S}_{t,i}^{(\mathcal{I})}) < \underbrace{\mathcal{S}_{expected,i}^{(\mathcal{I})}}_{ \Big(\mathbb{E}_{t}\Big(\frac{-\mathcal{P}_{c} \left(-erf\left(\frac{\mathcal{R'}}{\sqrt{2}\sigma_{R''}}\right) \right)}{2}
+ \lambda_{expected} \Big)\Big)},
\end{equation}
where stopping points are observable from  Mayer, Lagrange and Supremum (MLS-OST)~\cite{Peskir2006} for each CH at negligible Heston variance~\cite{benhamou2010time}, such that
\begin{equation}\label{eq:c2}
\fontsize{7}{8}\selectfont
  \mathcal{S}_{t,i}^{(\mathcal{I})}=\sup_{0\leq t \leq \tau} \mathbb{E}_{t}\left(\int_{0}^{\mathcal{R''}}f(p)^{(t)}.\mathcal{P}_{c}dp+\sup_{0\leq t \leq \tau} f(\mathcal{B'})\right).
\end{equation}
Using Lemma-1 and Lemma-2, (\ref{eq:c2}) deduces to \begingroup\makeatletter\def\f@size{7}\check@mathfonts$$
\max\Big(\mathbb{E}_{t}\Big(\frac{-\mathcal{P}_{c} \left(erf\left(\frac{\mathcal{R'}-\mathcal{R''}}{\sqrt{2}\sigma_{R''}}\right)-erf\left(\frac{\mathcal{R'}}{\sqrt{2}\sigma_{R''}}\right) \right)}{2}+ 2 \lambda_{1}\Big)\Big).~\refstepcounter{equation}(\theequation)\label{myeq4}$$\endgroup
\item If OST does not hold, Lemma-2 is used for changing CH, i.e. $\mathcal{S}_{t,i}^{(\mathcal{I})}=\underbrace{\mathcal{K}_{\mathcal{R},i}^{(\mathcal{T})}}_{\text{Upper limit of handling transactions}}$ and if $\mathcal{S}_{t,i}^{(\mathcal{I})} < \mathcal{K}_{\mathcal{R}}^{(\mathcal{T})}$, select CH that satisfies this condition provided that $\mathcal{R} \leq \mathcal{R''}$. If no CH is available, divide load based on $\max\left(\mathcal{R''}\right)$, or divide the load based on $\max\left(\mathcal{F}_{t}^{(\tau)}\right)$ for the given conditions in (\ref{eq:8}).
\item If both the above metrics fail to decide or create ambiguity, $\mathcal{S}_{t,i}^{(\mathcal{I})}$ is modeled for pre-energy decaying, i.e. change CH, if
     $\left(\int_{0}^{\tau-\omega} \left(\lambda\left(\beta^{(t)}_{\mathcal{D},i}\right)+\epsilon \sqrt{\mathcal{\sigma}} \frac{d\mathcal{B}'}{dt} \right)dt\right)_{expected}$ $<$
        \begingroup\makeatletter\def\f@size{6}\check@mathfonts$$
      \Bigg( \frac{2\lambda_{1}\mathcal{B}^{(o)}_{\mathcal{S}, i}}{|\mathcal{S}|}.$$\endgroup\begingroup\makeatletter\def\f@size{7}\check@mathfonts$$\dfrac{1-\left(\mathrm{e}^{-\left(\sqrt{2}\left(\sigma_2\sqrt{-\ln\left(\sqrt{2 \pi}f_2\sigma_2\right)}+\sigma_1\sqrt{-\ln\left(\sqrt{2 \pi}f_1\sigma_1\right)}\right)+\overline{\lambda_2}+\overline{\lambda_1}\right)(\tau-\omega)}\right)}{\sqrt{2}\left(\sigma_2\sqrt{-\ln\left(\sqrt{2 \pi}f_2\sigma_2\right)}+\sigma_1\sqrt{-\ln\left(\sqrt{2 \pi}f_1\sigma_1\right)}\right)+\overline{\lambda_2}+\overline{\lambda_1}}$$\endgroup
      \begingroup\makeatletter\def\f@size{7}\check@mathfonts$$+2\lambda_{1}\epsilon\sqrt{\sigma_{1}\sigma_{2}}(\tau-\omega)\Bigg),~\refstepcounter{equation}(\theequation)\label{myeq5}$$\endgroup where $\omega$ is the previous time slot/iteration for which the observational values are available.
\end{itemize}
The proposed optimal solution is convergent and guarantees uniqueness at $\mathcal{R}>0,\;\mathcal{R'}>0,\;\mathcal{R''}>0$ and $\tau > 0$. However, the uniqueness is subject to $\lambda_{1}$ and $\lambda_{2}$. As discussed earlier, for practical cases, $\lambda_{2}$ is unknown, because of which the solution converges w.r.t. $\lambda_{1}$ only. For extreme large values ($\mathcal{R}\rightarrow\infty,\;\mathcal{R'}\rightarrow\infty,\;\mathcal{R''}\rightarrow\infty$), the optimal observation is affected and there can be more than one solution for (\ref{myeq1})-(\ref{myeq4}), because of which the identification of CH becomes ambiguous and (\ref{myeq5}) is used as a solution. However, such a case is most unlikely to occur in practical scenarios. All the procedures defined in the system model can be accumulated, as shown in Algorithm~\ref{algo1}, which suggests when to change the CH and time slot in which pre-offloading should be performed to avoid delays and conserve the energy. The operational complexity of the algorithm depends on $\mathcal{C}$ and $\psi$. The distributed and parallel operations can further help to conserve the energy associated with the onboard processing on each vehicle.

\begin{algorithm}[!ht]
\fontsize{6}{8}\selectfont
\caption{CH Selection and offloading}
\label{algo1}
\begin{algorithmic}[1]
\State \textbf{Input}: Set network metrics and set idealistic values for $\mathcal{S}_{p,i}^{(\mathcal{I})}$
\State \textbf{Set}: Initial CHs and mark slots, $\omega$, $t$
\State \textbf{repeat}
\State Compute energy ratings for each vehicle and find $\mathcal{S}_{t,i}^{(\mathcal{I})}$
\State Change CH if $\mathcal{S}_{t,i}^{(\mathcal{I})} < \mathcal{S}_{p,i}^{(\mathcal{I})}$
\State \textbf{until:} $t\leq \tau$, (\ref{eq:7}) and (\ref{eq:8}) hold
\State \textbf{Output}: Offloading at $t-\omega$ for $\mathcal{S}_{t-\omega,i}^{(\mathcal{I})} < \mathcal{S}_{p,i}^{(\mathcal{I})}$ and new CH
\end{algorithmic}
\end{algorithm}

\begin{figure}[!ht]
  \centering
  \includegraphics[width=150px]{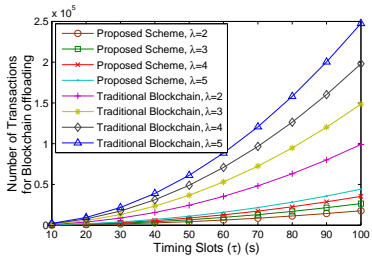}
  \caption{Number of Transactions vs. operational time.}\label{g1}
\end{figure}
\begin{figure}
  \centering
  \includegraphics[width=150px]{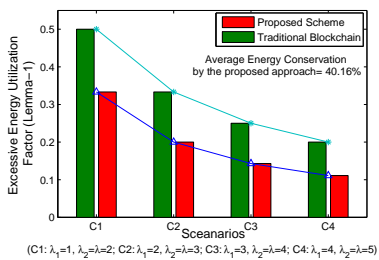}
  \caption{Energy conservation factor (Lemma-1) vs. requests per unit time.}\label{g2}
\end{figure}

\section{Performance Evaluations}
The proposed scheme was numerically evaluated for its efficacy compared with the traditional blockchain solution, which relies on the broadcasting of ledgers through Matlab\texttrademark. The evaluations were carried with settings $|\mathcal{C}|=5$, $|\mathcal{I}|=10$ in each cluster, $|\mathcal{S}|=10$, $\lambda=2\sim5$, $\gamma=\lambda$ requests per blockchain per second, $k=3$, $\lambda_1=1\sim4$, $\mathcal{H}=10$, $\lambda_2=2\sim5$, $\mathcal{E}_{R}=\mathcal{E}_{C}=2580$J, and $\beta_{C,M}^{(t)}=0.625$J is the approx. average, which can be fixed based on the curve algorithm or any dedicated security scheme~\cite{liu2008tinyecc}~\cite{potlapally2003analyzing}.  Excessive energy is taken twice as the total energy, i.e., $\epsilon=2$, $\tau=10\sim100$s, $\mathcal{R}=500$m~\cite{ozguner2011autonomous}, $\mathcal{R'}=\mathcal{R''}=300$m, $\omega=1$s, $\mathcal{P}_{c}=1$ i.e. maximum presence of vehicles. In addition, standard normal distribution was considered for normal and blockchain operations with zero mean and unit deviation. The proposed scheme helps to conserve energy by following a non-standard formation of clusters against the original policies of blockchain while keeping its principles intact. The performance was recorded for the number of transactions required to offload the entire blockchain (Fig.~\ref{g1}) and total energy conserved in variation with the number of requests (Fig.~\ref{g2}). The results in Fig.~\ref{g1} show that the proposed scheme overpowers the traditional blockchain by reducing the transaction load up to 82.06\% even at higher values for $\lambda$. These results govern the optimal solutions for selecting CH whenever the network's energy consumption goes beyond the set limits. Moreover, the proposed scheme shows an average energy conservation of 40.16\% compared with traditional blockchain at varying requests. Although the conserved energy decreases with an increase in the number of requests, the increase is lower than the standard operations and the conservation follows the values 33.32\%, 40.02\%, 42.85\%, 44.45\% for different values of $\lambda$, $\lambda_1$ and $\lambda_2$, as shown in Fig.~\ref{g2}. At present, the security of the network depends on the success of transactions and updates shared between the vehicles. The intermittent evaluations of the principles of the blockchain-security are beyond the scope of this article and will be discussed in future reports.
\section{Conclusion}
This letter aims at reducing the number of transactions required to update ledgers on the Internet of Vehicles. The purpose of the proposed approach is to reduce the burden of the network from numerous blockchain-transfer operations while conserving a maximum amount of available energy. To satisfy this goal, a distributed clustering model is designed which helps to conserve energy by 40.16\% (average) and reduces the number of ledger-transactions by 82.06\% compared with the traditional blockchain-based Internet of Vehicles (IoV).
\bibliographystyle{ieeetr}
\bibliography{refer_text}

\begin{thebibliography}{10}

\bibitem{gu2018consortium}
J.~Gu, B.~Sun, X.~Du, J.~Wang, Y.~Zhuang, and Z.~Wang, ``Consortium
  blockchain-based malware detection in mobile devices,'' {\em IEEE Access},
  vol.~6, pp.~12118--12128, 2018.

\bibitem{liu2018blockchain}
H.~Liu, Y.~Zhang, and T.~Yang, ``Blockchain-enabled security in electric
  vehicles cloud and edge computing,'' {\em IEEE Network}, vol.~32, no.~3,
  pp.~78--83, 2018.

\bibitem{uhlemann2018time}
E.~Uhlemann, ``Time for autonomous vehicles to connect [connected vehicles],''
  {\em IEEE Vehicular Technology Magazine}, vol.~13, no.~3, pp.~10--13, 2018.

\bibitem{wei2018qos}
C.-Y. Wei, A.~C.-S. Huang, C.-Y. Chen, and J.-Y. Chen, ``Qos-aware hybrid
  scheduling for geographical zone-based resource allocation in cellular
  vehicle-to-vehicle communications,'' {\em IEEE Communications Letters},
  vol.~22, no.~3, pp.~610--613, 2018.

\bibitem{lim2017preserving}
J.~Lim, H.~Yu, K.~Kim, M.~Kim, and S.-B. Lee, ``Preserving location privacy of
  connected vehicles with highly accurate location updates,'' {\em IEEE
  Communications Letters}, vol.~21, no.~3, pp.~540--543, 2017.

\bibitem{8326513}
X.~Liu, W.~Wang, D.~Niyato, N.~Zhao, and P.~Wang, ``Evolutionary game for
  mining pool selection in blockchain networks,'' {\em IEEE Wireless
  Communications Letters}, pp.~1--1, 2018.

\bibitem{benhamou2010time}
E.~Benhamou, E.~Gobet, and M.~Miri, ``Time dependent heston model,'' {\em SIAM
  Journal on Financial Mathematics}, vol.~1, no.~1, pp.~289--325, 2010.

\bibitem{Peskir2006}
G.~Peskir and A.~Shiryaev, {\em Stochastic processes: A brief review},
  pp.~53--121.
\newblock Basel: Birkh{\"a}user Basel, 2006.

\bibitem{liu2008tinyecc}
A.~Liu and P.~Ning, ``Tinyecc: A configurable library for elliptic curve
  cryptography in wireless sensor networks,'' in {\em International conference
  on Information processing in sensor networks}, pp.~245--256, IEEE, 2008.

\bibitem{potlapally2003analyzing}
N.~R. Potlapally, S.~Ravi, A.~Raghunathan, and N.~K. Jha, ``Analyzing the
  energy consumption of security protocols,'' in {\em International symposium
  on Low power electronics and design}, pp.~30--35, ACM, 2003.

\bibitem{ozguner2011autonomous}
U.~Ozguner, T.~Acarman, and K.~A. Redmill, {\em Autonomous ground vehicles}.
\newblock Artech House, 2011.

\end{thebibliography}
\end{document}